\documentclass[a4paper,11pt]{article}
\usepackage{jheppub} 
\usepackage{longtable} 
\usepackage{booktabs} 
\usepackage{lineno}
\usepackage{placeins} 
\usepackage{graphicx}
\usepackage{dcolumn}
\usepackage{bm}
\usepackage{amssymb}
\usepackage{amsmath}
\usepackage{xspace}
\usepackage{amsmath}
\usepackage{graphicx}
\usepackage{slashed}
\usepackage{lineno}
\usepackage{comment}
\usepackage{color}
\usepackage{soul}
\usepackage{setspace} 

\newcommand{\pp}           {pp\xspace}
\newcommand{\ppbar}        {\mbox{$\mathrm {p\overline{p}}$}\xspace}

\newcommand{\PbPb}         {\mbox{Pb--Pb}\xspace}

\newcommand{\s}            {\ensuremath{\sqrt{s}}\xspace}

\newcommand{\pt}           {\ensuremath{p_{\rm T}}\xspace}

\newcommand{\y}         {\ensuremath{y}\xspace}

\newcommand{\dsigmadpt}         {\ensuremath{\mathrm{d}\sigma/\mathrm{d}p_{\rm T}}\xspace}
\newcommand{\dsigmady}         {\ensuremath{\mathrm{d}\sigma/\mathrm{d}y}\xspace}

\newcommand{\ccbar}      {\ensuremath{\mathrm{c}\rm\overline{c}}\xspace}
\newcommand{\bbbar}      {\ensuremath{\mathrm{b}\rm\overline{b}}\xspace}

\newcommand{\nineH}        {$\sqrt{s}~=~0.9$~Te\kern-.1emV\xspace}
\newcommand{\seven}        {$\sqrt{s}~=~7$~Te\kern-.1emV\xspace}
\newcommand{\twoH}         {$\sqrt{s}~=~0.2$~Te\kern-.1emV\xspace}
\newcommand{\twosevensix}  {$\sqrt{s}~=~2.76$~Te\kern-.1emV\xspace}
\newcommand{\five}         {$\sqrt{s}~=~5.02$~Te\kern-.1emV\xspace}
\newcommand{\thirteen}         {$\sqrt{s}~=~13$~Te\kern-.1emV\xspace}
\newcommand{\twosevensixnn}{$\sqrt{s_{\mathrm{NN}}}~=~2.76$~Te\kern-.1emV\xspace}
\newcommand{\fivenn}       {$\sqrt{s_{\mathrm{NN}}}~=~5.02$~Te\kern-.1emV\xspace}

\newcommand{\GeVc}         {Ge\kern-.1emV/$c$\xspace}
\newcommand{\MeVc}         {Me\kern-.1emV/$c$\xspace}
\newcommand{\TeV}          {Te\kern-.1emV\xspace}
\newcommand{\GeV}          {Ge\kern-.1emV\xspace}
\newcommand{\MeV}          {Me\kern-.1emV\xspace}
\newcommand{\GeVmass}      {Ge\kern-.2emV/$c^2$\xspace}
\newcommand{\MeVmass}      {Me\kern-.2emV/$c^2$\xspace}

\newcommand{\dzero}        {\ensuremath{{\rm D}^{0}}\xspace}

\newcommand{\jpsi}         {\ensuremath{{\rm J}/\psi}\xspace}
\newcommand{\Bzero}       {\ensuremath{\rm B^{0}\xspace}}   
\newcommand{\Bpm}       {\ensuremath{\rm B^{\pm}\xspace}}   
\newcommand{\Bs}       {\ensuremath{\rm B_{s}^{0}\xspace}}   
\newcommand{\Lb}       {\ensuremath{\rm \Lambda_{b}\xspace}}

\title{Data-driven analysis of the beauty hadron production in \pp collisions at the LHC with Bayesian unfolding}

\author[a,b]{Xiaozhi Bai}
\author[a,b]{Guangsheng Li}
\author[a,b,1]{Yifei Zhang,\note{Corresponding author.}}
\author[b]{Qingyi Situ}
\author[a,b]{Xiaolong Chen}

\emailAdd{xiaozhi.bai@cern.ch} 
\emailAdd{ego2017@mail.ustc.edu.cn} 
\emailAdd{ephy@ustc.edu.cn}

\affiliation[a]{State Key Laboratory of Particle Detection and Electronics}
\affiliation[b]{University of Science and Technology of China, Hefei 230026, China}

\abstract{Heavy flavour production in proton-proton (\pp) collisions provides insights into the fundamental properties of Quantum Chromodynamics (QCD). Beauty hadron production measurements are widely performed through indirect approaches based on their inclusive decay modes. A Bayesian unfolding data-driven analysis of the ALICE and LHCb data was performed in this study, which recovers the full kinematic information of the beauty hadrons via different inclusive decay channels. The corresponding beauty hadron production cross sections obtained after the Bayesian unfolding are found to be consistent within their uncertainties. The weighted average open beauty production cross sections are presented as a function of the transverse momentum and rapidity in \pp collisions at \s = 5.02 \TeV and \s = 13 \TeV, respectively. The \pt-integrated open beauty production \dsigmady and the total \bbbar cross section $\sigma_{\rm \bbbar}$ are also reported. The precision of these results significantly improves upon worldwide measurements, providing valuable validation and constraints on mechanisms of heavy flavour production in \pp collisions at the LHC energies.}

\begin{document}
\maketitle
\flushbottom

\section{Introduction}\label{sec:introduction} 

The production of heavy flavour (charm and beauty) plays a crucial role in both \pp and heavy-ion collisions. The study of heavy flavour production is important to understand the fundamental theory of Quantum Chromodynamics (QCD). The masses of the heavy quarks are significantly larger than the typical QCD energy scale, therefore, heavy quark production cross section in \pp collisions can be evaluated by perturbative QCD (pQCD) calculations~\cite{Cacciari:2005rk,Frixione:2005yf}. Precise measurements of heavy-flavour hadron production cross section are important to validate pQCD theory in \pp collisions. Furthermore, the masses of the heavy flavour quarks are much larger than the typical temperature of the quark-gluon plasma (QGP), which is suggested to be created in heavy-ion collisions~\cite{Catani:2020kkl, Kramer:2017gct}. They are predominantly produced by hard partonic scattering processes in the early stage of the heavy-ion collisions and thus experience the full evolution of the system~\cite{Shuryak:2014zxa,Braun-Munzinger:2015hba,Busza:2018rrf,Norman:2023chk,ALICE:2013olq}. Therefore, the heavy flavour production measurements in \pp collisions are essential to test pQCD calculations and provide a reference for studying the nuclear medium effects and the properties of the QGP in heavy-ion collisions.  

The calculation of heavy flavour production in \pp collisions according to QCD can be factorized into several independent terms~\cite{cacciari2012theoretical,maciula2013open}: \romannumeral1) The Parton Distribution Functions (PDFs), describing the probability of the quark or gluon to carry a given fraction of the momentum of the incoming protons~\cite{Lai:2010vv,Ball:2017nwa} provide crucial inputs for the calculations and need to be constrained by experimental measurements. \romannumeral2) The hard scattering cross sections refer to the probability to produce a heavy quark pair via a specific process, such as a gluon-gluon interaction. These probabilities can be calculated by pQCD~\cite{cacciari2012theoretical,MAZZITELLI2023137991}. \romannumeral3) The produced heavy flavour partons fragment into particular hadrons according to the so-called Fragmentation Functions (FFs). The FFs are a typical non-perturbative process and need to be constrained by measurements~\cite{Metz:2016swz}. Therefore, the calculation of heavy flavour hadron production cross sections involves both the perturbative and non-perturbative aspects of QCD. The heavy quark production cross section is not only critical for verifying the framework for QCD calculations but also provide inputs for its calculation~\cite{Cacciari:1998it}.

Several experimental measurements of the heavy flavour production are carried out in \pp collisions from \s= 0.2 to 13 \TeV by the STAR, PHENIX, LHCb, CMS, ALICE and ATLAS collaborations~\cite{STAR:2021uzu,PHENIX:2019pxh,Aaij:2013noa,Aaij:2015fea,Aaij:2016avz,Khachatryan:2016csy,ALICE:2017olh,ATLAS:2021xtw}. The direct reconstruction of open beauty hadrons is challenged by the large combinatorial background and low specific hadronic decay branching ratios. An indirect approach, through inclusive decay modes is commonly employed in experimental measurements, which relies on the decay of open beauty hadrons to specific products and channels. Non-prompt components, originating from open beauty hadron decays, are frequently utilized in experimental measurements to assess beauty production~\cite{ALICE:2023hou,ALICE:2022tji}. For example, commonly studied decay channels of open beauty hadrons include non-prompt D-mesons, \jpsi, as well as single electrons and muons from heavy flavour decays~\cite{ALICE:2024xln,ALICE:2023hou,ALICE:2022iba,ATLAS:2019xqc,Si:2019esg,Li:2021ddz}. Given the very different masses of the corresponding decay products and their distinct decay kinematics, the measured non-prompt components \pt-differential distributions are impacted by the decay process and can not directly reflect the beauty hadrons production. Consequently, different measured results are often discussed individually for specific decay channels, despite originating from the same underlying beauty production and merely representing different decay channels.

The Bayesian Unfolding has been employed since 1994, initially introduced by G. D'Agostini. It is a statistical technique used to recover the entire information based on the partial experimental data via the statistical approach, namely the semi-measurements~\cite{2010arXiv1010.0632D}. This method is grounded in Bayesian statistical theory, which interprets probability as a measure of certainty, given prior theoretical inputs of the typically expected distribution~\cite{2004physics3086D,2003RPPh66.1383D}. Bayesian unfolding offers a framework that explicitly incorporates uncertainties in both the measurements and the prior inputs. One of the key features is its ability to provide a systematic way of evaluating the uncertainties associated with the unfolding process itself, including both statistical and systematic uncertainties~\cite{2010arXiv1010.0632D}.

In this article, the systematic study of the total beauty production cross section is carried out through the data-driven analysis via Bayesian unfolding. The new results of the total beauty production cross section as a function of the \pt, rapidity~\y, and the centre-of-mass energy are reported. The unfolded results from different experimental measured decay channels, namely non-prompt \dzero at midrapidity ($\lvert\y\lvert$ $<$ 0.5) and \jpsi at forward rapidity (2.0 $<$ \y $<$ 4.5), are found to be perfectly compatible within their uncertainties. The weighted averaged results are computed 
with the weights evaluated according to the uncorrelated uncertainties of the considered inclusive decay modes mentioned above. The total \bbbar cross section is evaluated by integrating over \pt and \y. The precision of this new approach is improved significantly compared with the previous ALICE published measurements via its individual decay channel~\cite{ALICE:2024xln,ALICE:2022iba}. 

The article is organized as follows: Section~\ref{sec:BayesianUnfold} provides the detailed description of the Bayesian unfolding statistical analysis method. A brief discussion of the uncertainties in this statistical analysis method is provided in Section ~\ref{SystmaticalUncertantanties}. The results of the total beauty production cross sections obtained through the Bayesian unfolding method, along with comparisons to previously published results from various measurements and theoretical pQCD calculations are presented in Section ~\ref{sec:results}. The discussions and conclusions from the results and the implications for the underlying physics are summarized in Section ~\ref{Conclusions}.

\FloatBarrier
\section{Bayesian unfolding statistical analysis method}\label{sec:BayesianUnfold} 

Bayesian unfolding is a statistical analysis technique used to infer the entire distribution based on partial information. In this article, a data-driven approach is used to carry out the inclusive measurements of open beauty hadron production via non-prompt \dzero and \jpsi, i.e., the semi-inclusive decay of \Bzero, \Bpm, \Bs, \Lb $\rightarrow$ \jpsi+X and \Bzero, \Bpm, \Bs, \Lb $\rightarrow$ \dzero+X~\cite{ALICE:2024xln,ALICE:2021mgk,LHCb:2021pyk,LHCb:2015foc}.  The decay products X are not used in the beauty hadron reconstruction. However, the missing information from the X component can be included in the simulation in the response matrix between the kinematic variables of the open beauty hadron and its decay products. Bayesian unfolding is a powerful technique for recovering the open beauty hadron entire information from its decay products. The essential idea is presented as follows in Eq.~\ref{eq:unfoedprobaliblity}, the Bayes' theorem allows to invert the response from A to B to a response from B to A, the latter being better known than the former. The $P(A \mid B)$ is the probability of $A$ given $B, P(B \mid A)$ is the probability of $B$ given $A, P(A)$ is the prior probability of $A$, and $P(B)$ is the prior probability of $B$.

\begin{equation}\label{eq:unfoedprobaliblity} 
P(A \mid B)=\frac{P(B \mid A) \cdot P(A)}{P(B)}
\end{equation}

The concept of this approach can be elucidated by the Eq.~\ref{eq:Bayesianunfolding}, where $\rm f_{M}(x)$ and $\rm f_{T}(y)$ represent the measured non-prompt products and open beauty hadron distributions of the quantities $x$ and $y$, respectively. The correspondence between quantities $x$ and $y$ is described by a response matrix  $R(x|y)$, which transforms the original distribution $\rm f_{T}(y)$ into the measured distribution $\rm f_{M}(x)$. 

\begin{equation}\label{eq:Bayesianunfolding} 
	\rm f_{M}(x) = \int R(x|y)f_{T}(y) \mathrm{d}y
\end{equation}
 
The realistic measured data are sorted into various bins such as \pt, \y, or multiplicity intervals. The original (or expected) distributions are represented as $\rm E_{i} (i=1,2,...n)$. Subsequently, after accounting for the correspondence between measurements and expectations through unfolding techniques utilizing the transformation $\rm U_{ij}$, Eq.~\ref{eq:Bayesianunfolding} follows a summation form, as shown in Eq.~\ref{eq:BayesianunfoldingSum}. Consequently, constraints for each bin interval $\rm (E_{i})$, and for a given effect like the open beauty hadron decay \pt smearing, can be generated from different bins $\rm M_{j} (j=1,2,...m)$ via migration.  $\rm P(M_{j}|E_{i})$ can represent the efficiency of reconstruction, decay response, or any other unknown effects, but it should be understood through simulation.
 
\begin{equation}\label{eq:BayesianunfoldingSum} 
\rm	E_{i} = \sum\limits^{m}_{j=1} U_{ij} M_{j}
	\end{equation}

\begin{equation}\label{eq:BayesianunfoldingRij} 
\rm	U_{ij} = P(E_{i} | M_{j}) / \epsilon_{i} =P(M_{j}|E_{i})\cdot P(E_{i})/P(M_{j})/ \epsilon_{i},
\end{equation}

\begin{equation}\label{eq:efficiency} 
\rm	\epsilon_{i} = \sum_{j=1}^{m}P(M_{j}|E_{i})
\end{equation}
 
\begin{figure}[!htb]
	\begin{center}
		\includegraphics[width = 14cm]{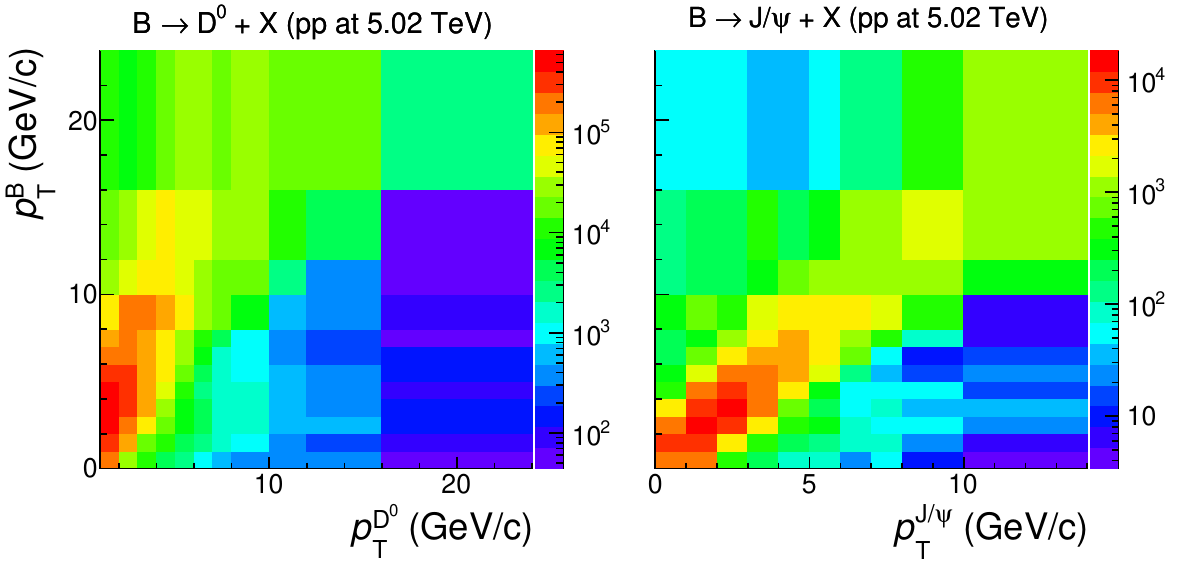}
              \includegraphics[width = 14cm]{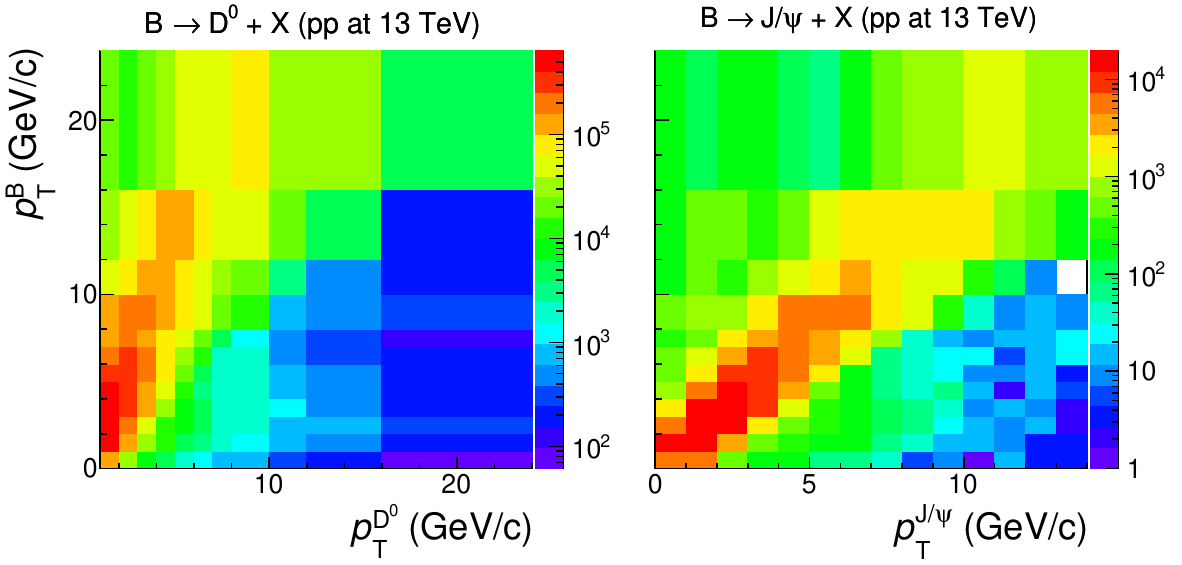}
		\caption{Response matrix of the beauty hadrons as a function of the non-prompt \dzero (left panels) and \jpsi (right panels) in \pp collisions at \s = 5.02 (upper panels) and 13 \TeV (lower panels) at rapidity interval $\lvert$y$\lvert$ $<$ 10.}   
		\label{fig:unford-matrix}
	\end{center}
\end{figure}

The expression for $\rm U_{ij}$ is explained by Eq.~\ref{eq:BayesianunfoldingRij}, and the details have been discussed in the previous equation~\ref{eq:unfoedprobaliblity}. The response matrix delineates the correlation between measured data and the expected values of beauty hadrons. It is derived through  EvtGen 2.0~\cite{LANGE2001152} simulations, EvtGen is a Monte Carlo event generator specifically designed for simulating the decays of heavy flavour hadrons, particularly beauty hadron decays. Additionally, the energy loss in the final state radiation is considered and simulated by the PHOTOS 3.64 package~\cite{BARBERIO1991115}. All the detected efficiency including the decay acceptance, and decay branching ratio effects for observing non-prompt \dzero and non-prompt \jpsi, can be described by $\rm \epsilon_{i}$, which is defined in Eq.~\ref{eq:efficiency}. The illustration of the response matrix of the beauty hadrons decays into \dzero and \jpsi are shown in \pp collisions at \s = 5.02 (upper) and 13 (lower) \TeV on the Fig.~\ref{fig:unford-matrix}, in the left and right panels, respectively. 

Equation~\ref{eq:BayesianunfoldingSum} and ~\ref{eq:BayesianunfoldingRij} need the knowledge of the beauty hadron original distribution $\rm E_{i}$, which is not known, so it is necessary to start from a prior distribution from FONLL calculations~\cite{Cacciari:1998it,Cacciari:2015fta} , and then through the iterative method to have a better estimation of it, as described in Ref.~\cite{DAGOSTINI1995487}. Uncertainties are evaluated and systematically propagated, with the cumulative effect of all iterations taken into account. The iteration is halted once the difference between the current and previous iterations is significantly smaller than the uncertainties.

\FloatBarrier
\section{Uncertainties analysis}
\label{SystmaticalUncertantanties} 

The sources of uncertainty considered in this study are dominated by the uncertainties of the measured non-prompt \dzero and non-prompt \jpsi, as well as semi-inclusive beauty hadrons (\Bzero, \Bpm, \Bs, \Lb) decay branching ratio to \dzero and \jpsi. Furthermore, the statistical uncertainties of the unfolding matrix, the number of Bayesian iterations, and the extrapolation of non-prompt \dzero \pt down to 0 are considered as the remaining systematical uncertainties. The evaluated total uncertainties are summarized in Table~\ref{tab:sys}, details are discussed in this section.

The main uncertainties originate from the measured data of non-prompt \dzero and \jpsi~\cite{ALICE:2024xln,ALICE:2021mgk,LHCb:2021pyk,LHCb:2015foc}. The non-prompt \dzero results provided by ALICE introduce relatively larger uncertainties compared to \jpsi, the latter measured by LHCb. The statistics and systematical uncertainties of measured non-prompt \dzero and \jpsi, are propagated separately to the final open beauty hadron production cross section. The statistical uncertainties can be considered independent over \pt, while the systematic uncertainties should be partially correlated among different \pt bins. The extent of this correlation is not straightforward to be determined. Therefore, two extreme assumptions are validated: fully correlated and independent. These correspond to the values of 4.3\% and 2.3\% for \s= 5.02 \TeV, and 5.2\% and 2.8\% for \s= 13 \TeV, respectively. The fully correlated case is considered by default, and results are reported in Table~\ref{tab:sys}. Therefore, the systematic uncertainties from the measured non-prompt \dzero and \jpsi are conservatively overestimated in this study.

The uncertainties of the inclusive beauty hadron decay to \dzero and \jpsi, are considered as global uncertainty, they are provided by the Particle Data Group (PDG)~\cite{pdf2024}. The statistical uncertainty of the response matrix in the unfolding process can be reduced by increasing the statistics of the EvtGen simulation, and it is below 1\% in this study. The uncertainty of the number of Bayesian iterations is evaluated by comparing the results of iterations 4 and 6, and the difference between the two cases is found to be much smaller than the statistical uncertainties. The discrepancy between the former and later iterations is around 1\%, which is considered as the systematical uncertainty of the number of iterations. 
 
The minimum \pt of the measured non-prompt \dzero by ALICE is 1 \GeVc, while the measured non-prompt \jpsi by LHCb extends down to 0 \GeVc. Ideally, no extrapolation is needed of the \pt, as the contribution of unmeasured \pt and \y intervals should be recovered by the unfolding process.  The difference in results with and without the extrapolation is lower than 1\%. The difference between with and without extrapolation is considered as the uncertainty of the measured non-prompt \dzero \pt down to 0. 

The uncertainty on the beauty hadron \pt shape for the realistic response matrix creation is considered as the last component of the systematical uncertainties. This is evaluated via re-weighting the \pt shape of the open beauty hadron used for the response matrix creation, using two different \pt shapes: one from PYTHIA 8.309 and another from the FONLL prediction. The corresponding discrepancy observed on the final results is around 1\% for all cases, except for non-prompt \jpsi at \s= 13 \TeV, where it is 5.2\%.
 
\begin{figure}[!h]
\begin{center}
 \includegraphics[width = 7cm]{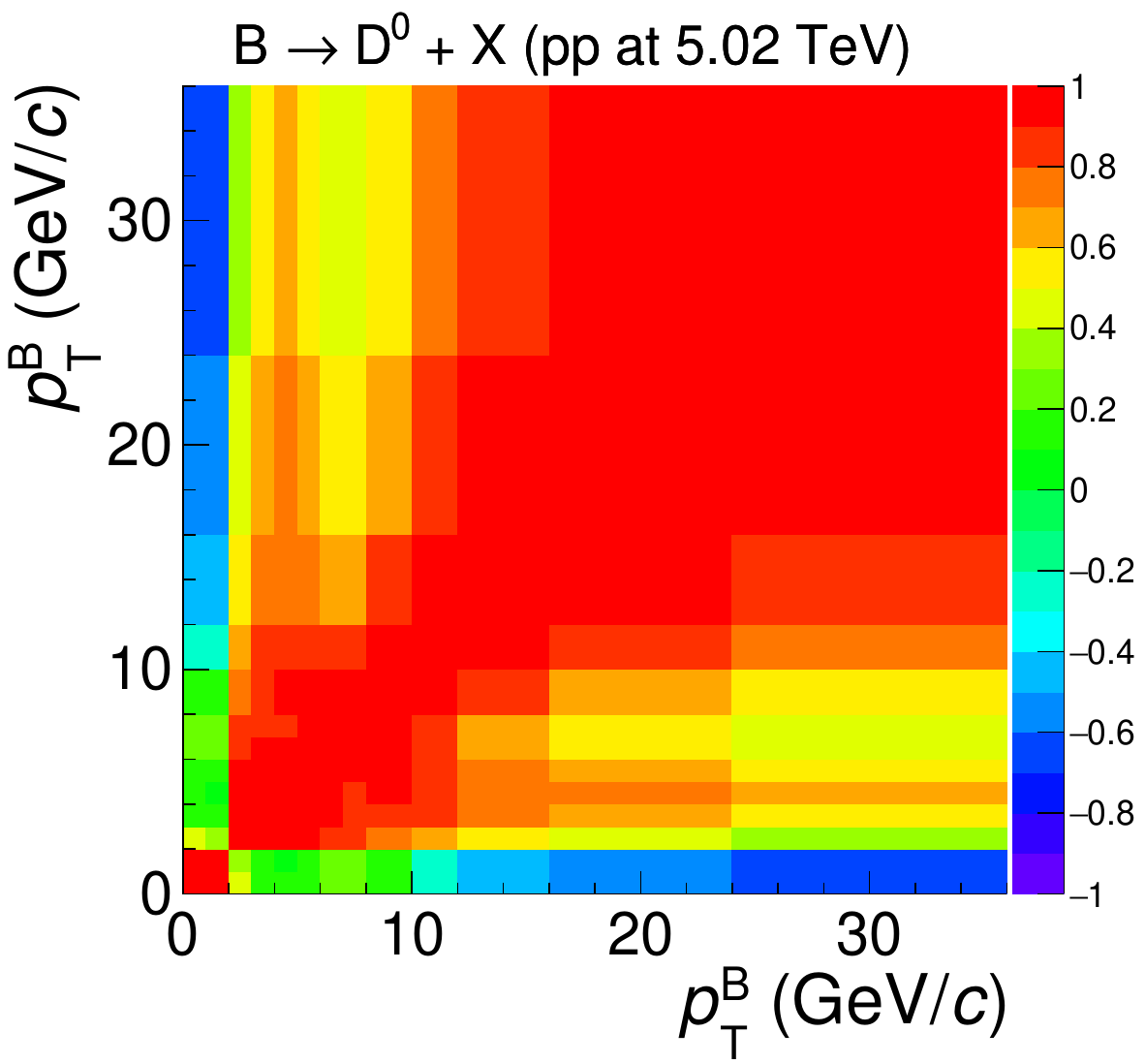}
 \includegraphics[width = 7cm]{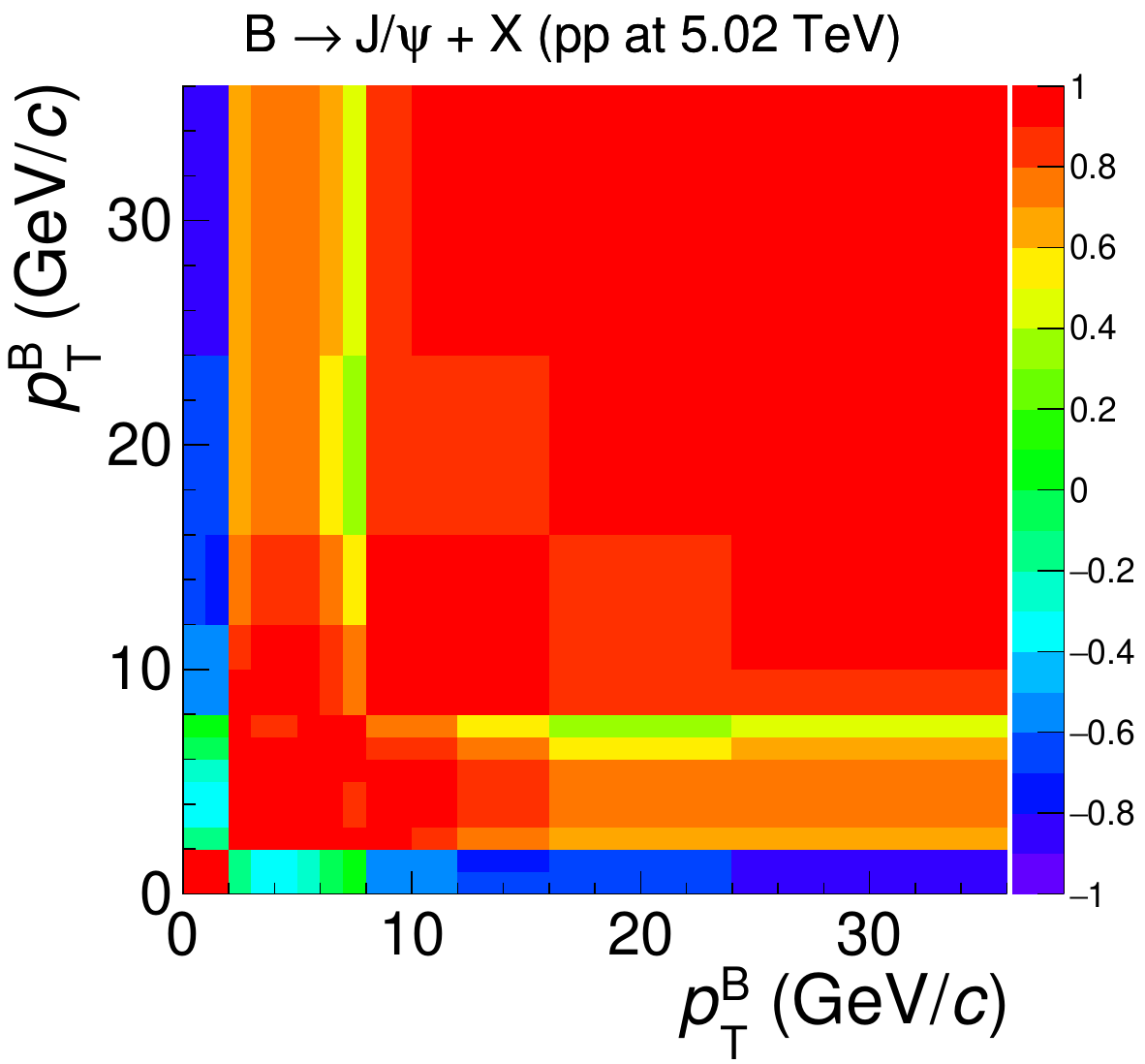}
 \includegraphics[width = 7cm]{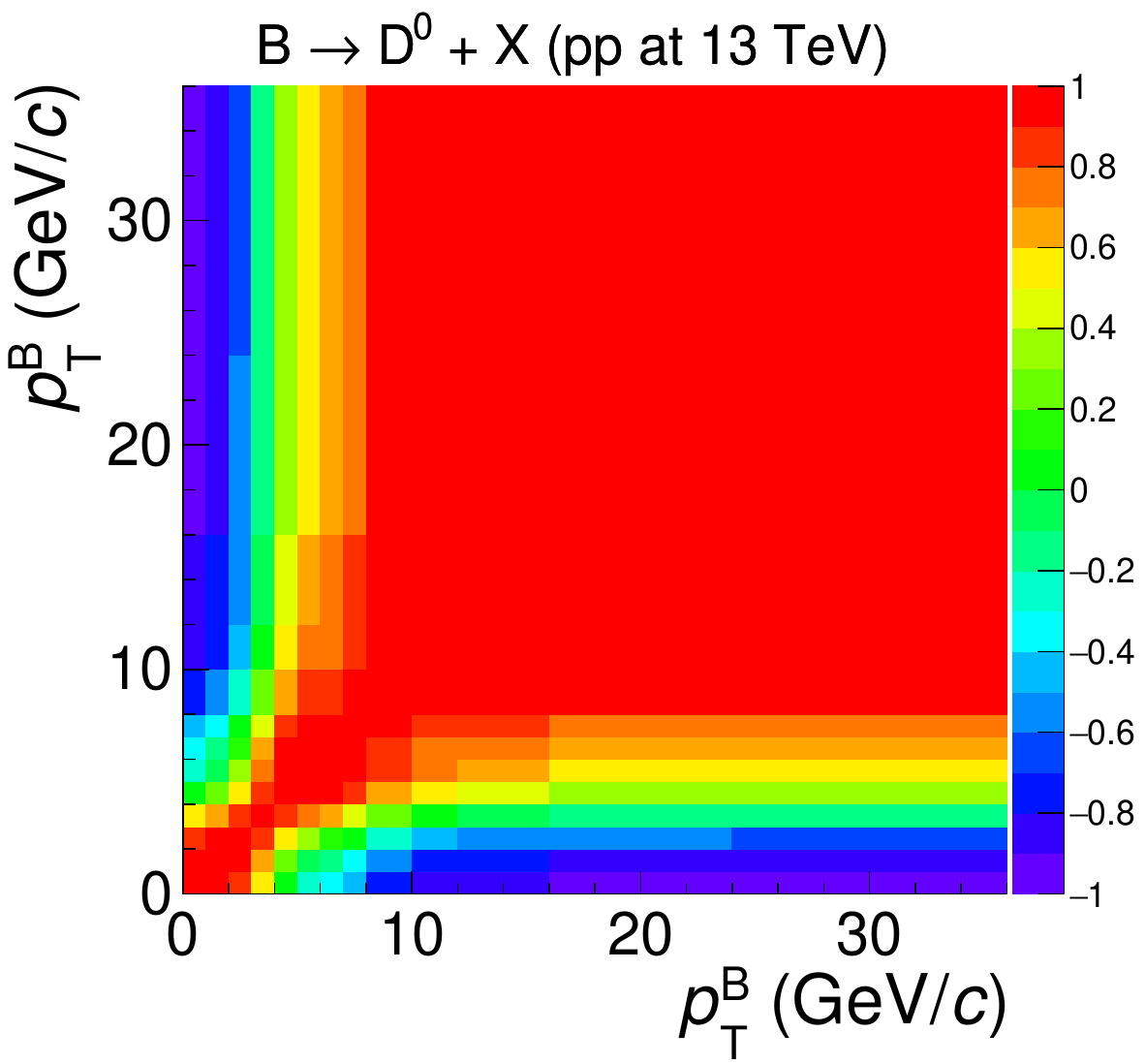}
 \includegraphics[width = 7cm]{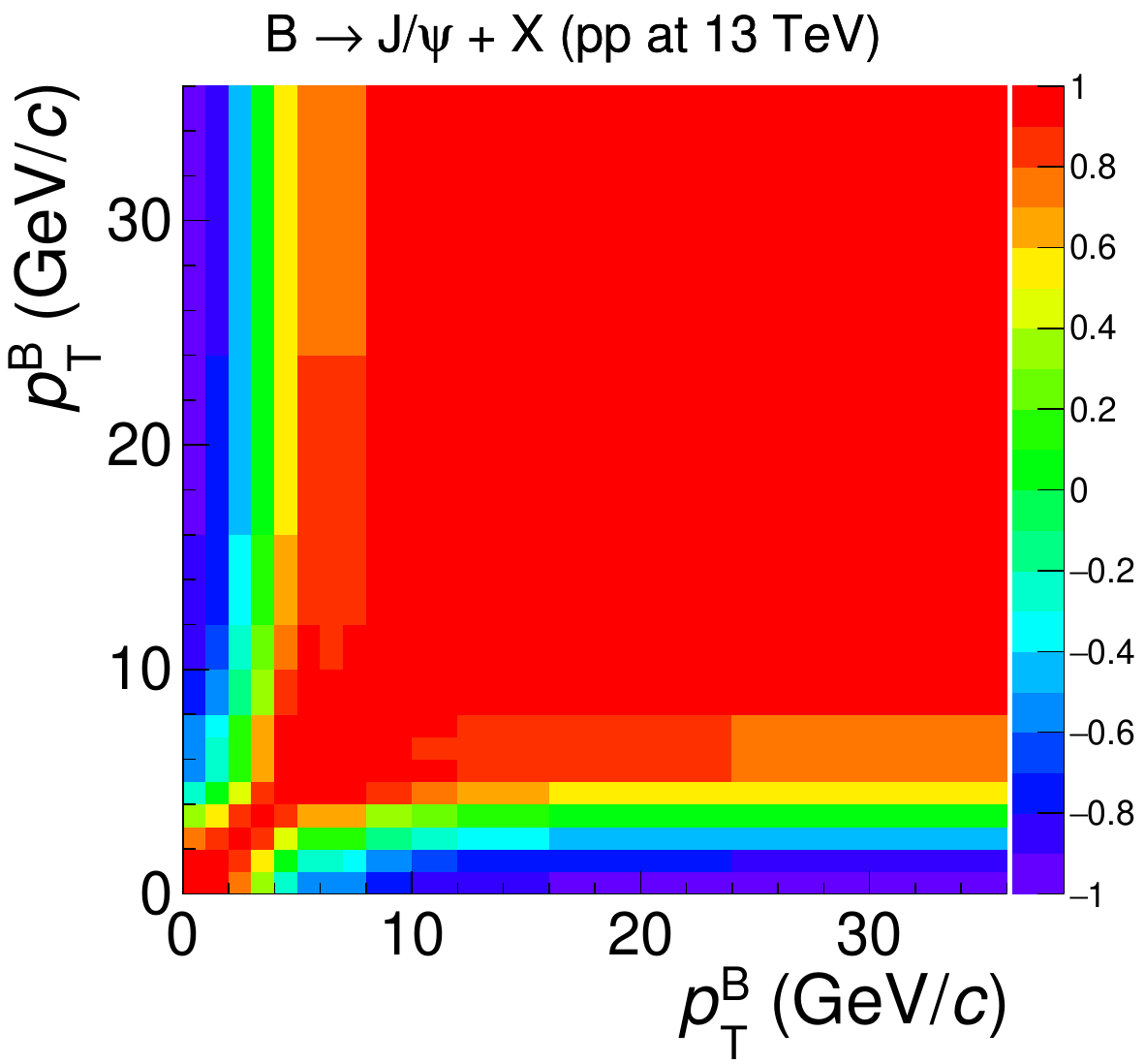}
\caption{Covariance matrix of \bbbar production cross section from the unfolded non-prompt \dzero (left panels), and \jpsi (right panels) in \pp collisions at \s = 5.02 (upper panels) and 13 \TeV at rapidity interval $\lvert$y$\lvert$ $<$ 10.}
\label{fig:CovarianceMetrix}
\end{center}
\end{figure}

Different sources of uncertainties are described, the correlated and uncorrelated uncertainties are evaluated and propagated separately, and the corresponding values are listed in Table~\ref{tab:sys}. The final uncertainties are summed up in quadrature. The weighted average total uncertainties for \bbbar production cross section at midrapidity are 8.2\% and 8.1\% for \s= 5.02 \TeV and  \s= 13 \TeV, respectively. Fig.~\ref{fig:CovarianceMetrix} shows the relative covariance matrix, also known as the correlation matrix, normalizing the covariances by the standard deviations of the respective variables, which provides the strength and direction of the relationship between the unfolded results at different \pt intervals from the decay of non-prompt \dzero or \jpsi. The diagonal elements of the relative covariance matrix are always 1, as they represent the correlation of each variable with itself. The off-diagonal elements range from -1 to 1, where 1 indicates a perfect positive correlation, -1 indicates a perfect negative correlation, and 0 indicates no correlation.

\begin{table*}[!h]
	\caption{Systematic uncertainties on the \pt-integrated \bbbar cross section, obtained from the Bayesian unfolding in \pp collisions at \s = 5.02 and \s = 13 \TeV at midrapidity, are reported. The individual contributions and the total uncertainties are given in percentage. It is considered that all the systematical uncertainties are correlated over \pt.} 
		\begin{tabular}{c|ccc|ccc}
		   \hline
			Uncertainty sources & \multicolumn{3}{c|}{\five} & \multicolumn{3}{c}{\thirteen} \\
			\hline
			& $\dzero$ & $\jpsi$ & average & $\dzero$ & $\jpsi$ & average \\
			\hline
			Measured non-prompt data stat. & $9.1$  & $0.4$ & $2.6$  & $4.2$ & $0.5$ & $2.1$ \\
			\hline
			Measured non-prompt data syst. & $10.5$  & $4.4$ & $4.3$  & $8.4$ & $6.2$ & $5.2$ \\
			\hline
			Response matrix stat.      & $<0.1$  & $0.7$ & $0.5$  & $<0.1$ & $0.3$ & $0.1$ \\
			\hline
			Branching ratio of $\rm h_{B}$ decay           & $5.8$  & $8.6$ & $6.4$  & $5.8$ & $8.6$ & $5.2$ \\
			\hline
			Bayes iterations         & $0.3$  & $1.4$ & $1.0$  & $0.6$ & $<0.1$ & $0.3$ \\
			\hline
			W/o \pt extrapolation            & $0.1$ & N/A     & $<0.1$ & $0.4$ & N/A & $0.2$ \\
			\hline
			Beauty hadron \pt shape & $1.0$  & $0.8$ & $0.6$  & $1.6$ & $5.2$ & $2.7$ \\
			\hline
			\hline
			Total uncertainty & $15.2$  & $9.8$ & $8.2$  & $11.2$ & $11.8$ & $8.1$ \\
		   \hline
		\end{tabular}
	\label{tab:sys}
\end{table*}

\FloatBarrier

\section{Results}\label{sec:results} 
In this section, \pt-differential (\dsigmadpt) as well as \y-differential (\dsigmady, \pt $>$0) cross section results of \bbbar production are provided in \pp collisions at \s = 5.02 and 13 TeV. These results are obtained from the Bayesian unfolding procedure applied to the measured non-prompt \dzero and non-prompt \jpsi cross section measurements at the same energies. The unfolded results are corrected by the inclusive beauty hadrons (\Bzero, \Bpm, \Bs, \Lb) decay to \jpsi and \dzero branching ratio from PDG~\cite{pdf2024}, the values are 60.4 $\pm$ 3.5\%  and  1.16 $\pm$ 0.10\% for  \dzero and \jpsi, respectively.

\begin{figure}[!htb]
\begin{center}
 \includegraphics[width = 7.5cm]{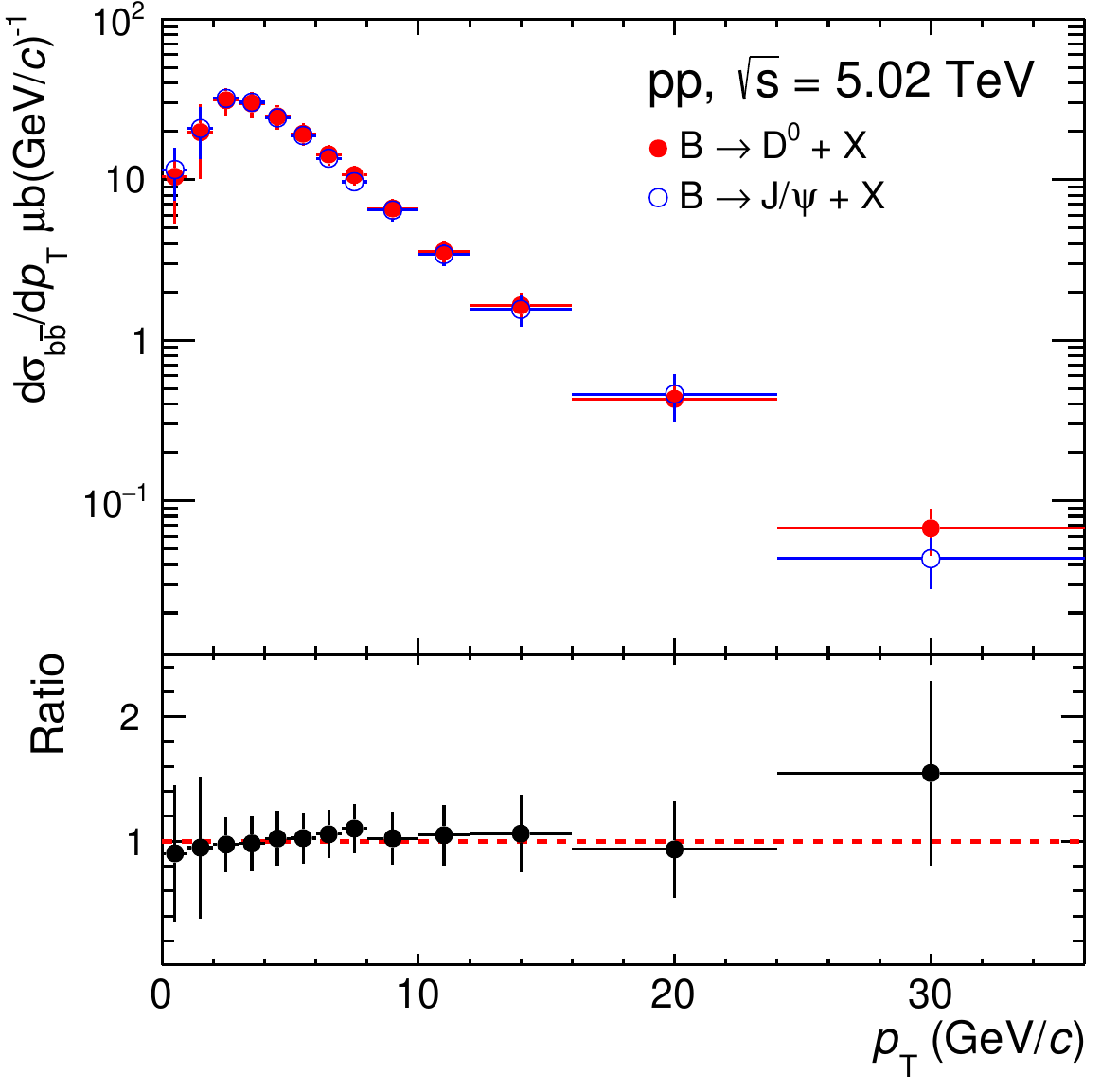}
 \includegraphics[width = 7.5cm]{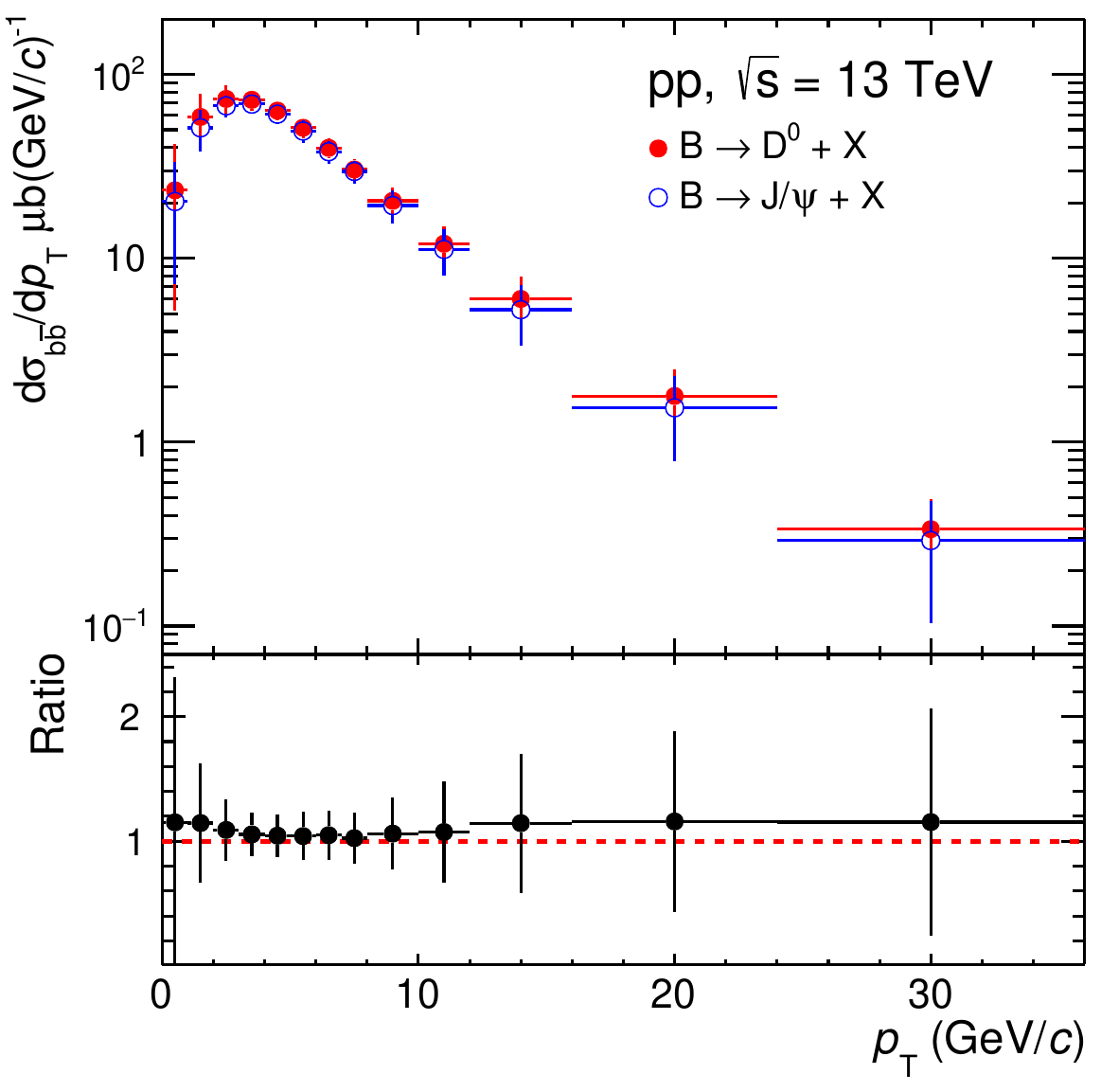}
\caption{The \bbbar production cross section \dsigmadpt as a function of \pt in \pp collisions at \s = 5.02 \TeV (left panel) and \s = 13 \TeV (right panel) at rapidity interval $\lvert$y$\lvert$ $<$ 10, the data points represent the unfolded results from the decay channels to non-prompt \dzero(red) and \jpsi(blue).}
\label{fig:JpsiPtSpectra_three_5020}
\end{center}
\end{figure}

The critical inputs of unfolding analysis are the measured non-prompt components from the open beauty hadron decays. The measured non-prompt \dzero from ALICE~\cite{ALICE:2021mgk} and non-prompt \jpsi~\cite{LHCb:2021pyk} from LHCb are used in \pp collisions at \s = 5.02 \TeV analysis, while the similar inputs in \pp collisions at \s = 13 \TeV can be found in Refs.~\cite{ALICE:2024xln,LHCb:2015foc}. The \pt-differential production cross section \dsigmadpt via Bayesian unfolding analysis is shown at rapidity interval $\lvert$y$\lvert$ $<$ 10 in Fig.~\ref{fig:JpsiPtSpectra_three_5020}. The data points represent the results obtained by unfolding from the two aforementioned inclusive decay channel measurements of non-prompt \dzero (red) and non-prompt \jpsi (blue). The results of \bbbar production cross section \dsigmadpt as a function of the \pt in \pp collision at \s = 5.02 \TeV and \s = 13 \TeV are shown in the left and right panels, respectively. The results derived from two different semi-inclusive decay channels (non-prompt \dzero and non-prompt \jpsi) are compared in the two upper panels, and the corresponding ratios are shown in two lower panels. It is found that different semi-inclusive decay channels are perfectly consistent within their uncertainties after unfolding the beauty hadron to the same rapidity intervals ($\lvert$y$\lvert$ $<$ 10), although the measured non-prompt \dzero and non-prompt \jpsi are from very different rapidity coverage. The non-prompt \dzero is measured at midrapidity ($\lvert$y$\lvert$ $<$ 0.5), while the non-prompt \jpsi measurements are at forward rapidity (2.5 $<$ y $<$ 4), the large rapidity gap is recovered by the unfolding matrix. This is expected, as they originate from the decay of the same underlying beauty hadrons once they are unfolded to identical kinematic ranges. The consistency indicates the unfolding matrix produced by the EvtGen describes properly the dependence of open beauty hadron and its decay products. Small differences among different decay channels could result from statistical fluctuations in the measured non-prompt \dzero and \jpsi. Furthermore, the consistences can provide an important cross check among the measurements provided by different collaborations.

It is natural to compute the weighted average \pt-differential \bbbar production cross section on results obtained from different decay channels, according to the Eq.~\ref{eq:averageD}. The inversed quadratic uncertainties are used as the weight factor in the computation of the averaged \pt-differential beauty hadron production cross sections, as shown in Eq.~\ref{eq:averageDuncer}. This method has been utilized by ALICE for computing the averaged charm mesons production yields in \PbPb collision at \fivenn~\cite{ALICE:2021rxa}. The main idea behind this method is that data points with smaller uncertainties carry a relatively larger weight factor in combined results calculations. 
 
 \begin{equation}\label{eq:averageD} 
 \dsigmadpt(\rm {ave.})=\sum_{i=(\dzero,\jpsi)}w(i) \dsigmadpt(i)
 \end{equation}
 
  \begin{equation}\label{eq:averageDuncer} 
 \rm w(i)  = \frac{1/\sigma^{2}_{i}} {\sum_{ i=(\dzero,\jpsi)} 1/\sigma^{2}_{i}}
 \end{equation}
  
 \begin{figure}[!htb]
 	\begin{center}
 		\includegraphics[width = 7.5cm]{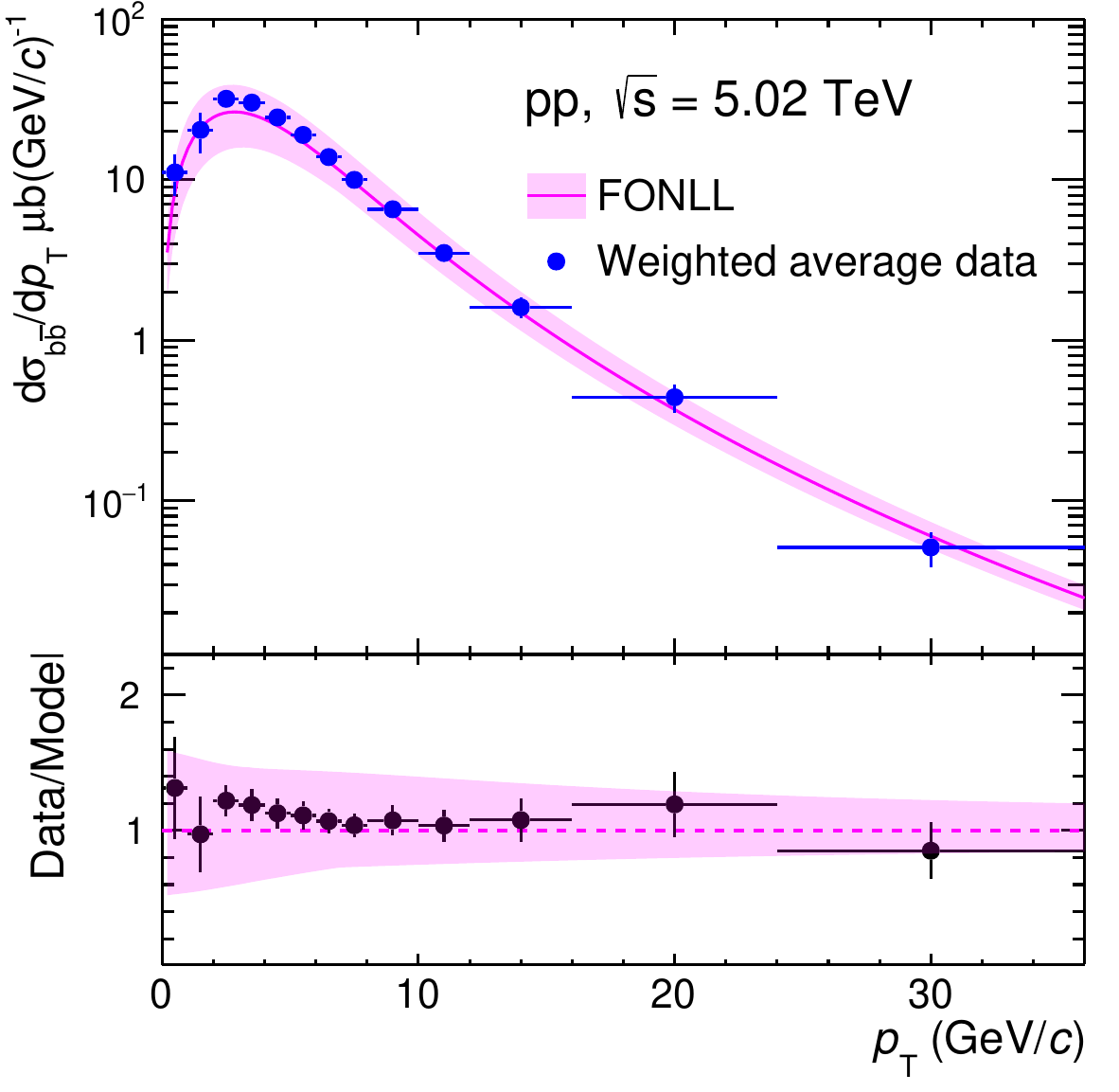}
            \includegraphics[width = 7.5cm]{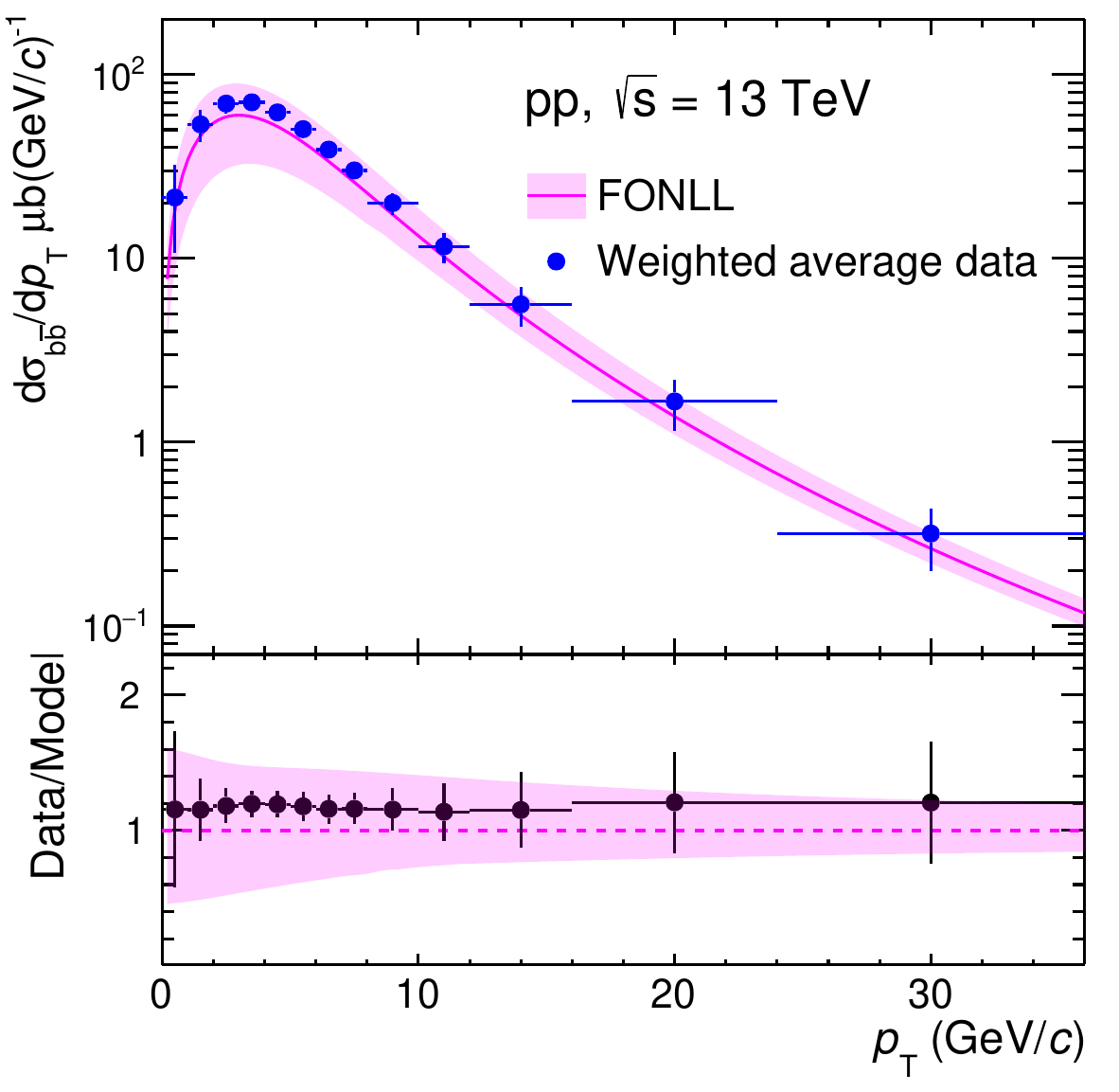}
 		\caption{The comparison of the weight averaged \bbbar production cross section and the FONLL calculations as a function of \pt in \pp collisions at \s = 5.02 \TeV (left panel)and \s = 13 \TeV (right panel), the ratios of the \bbbar production cross section to FONLL calculations are shown in the two lower panels. }
 		\label{fig:JpsiPtSpectra_three_5020-ratio}
 	\end{center}
 \end{figure}

The \pt-integrated \bbbar production cross section per unit of rapidity \dsigmady in \pp collisions at \s = 5.02 \TeV, is calculated by the integral of the weighted average \pt-differential cross section \dsigmadpt. The \pt-integrated uncertainties need to consider the correlation over \pt. The bin immigration is simulated by the unfolding, and the correlations between different bins are described by the covariance matrix, which is provided alongside with the unfolding matrix. 

\begin{figure}[!htb]
\begin{center}
 \includegraphics[width = 14cm]{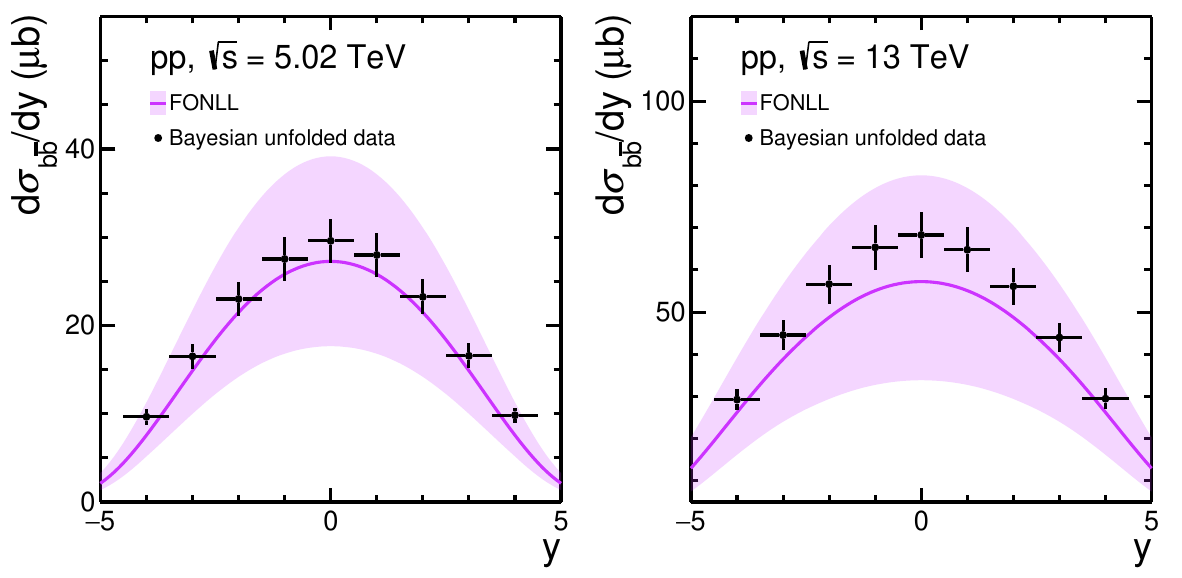}
\caption{The comparison of the weighted average \pt-integrated \bbbar production cross section and the FONLL calculations as a function of rapidity(\y) in \pp collisions at \s = 5.02 \TeV (left panel) and \s = 13 \TeV(right panel).}  
\label{fig:bbarrapidty}
\end{center}
\end{figure}

The results of the total \bbbar cross section as a function of the rapidity in \pp collisions at \s = 5.02 \TeV are shown in Fig.~\ref{fig:bbarrapidty} left panel. The \pt-integrated \bbbar production cross section is compared with the FONLL predictions. The FONLL calculation can describe the unfolded results within the rapidity interval $\lvert$y$\lvert$ $<$ 4.5, despite its large uncertainties in the calculations. To validate the FONLL calculations, the uncertainties of the calculation need to be improved. It is dominated by the PDFs for the low-\pt and large-\y regions, and renormalisation and factorisation scales dominate the high-\pt and small-\y regions.

The \pt-integrated \bbbar production cross section per unit of rapidity \dsigmady in \pp collisions at \s = 13 \TeV is obtained using the same procedures as in \pp collisions at \s = 5.02 \TeV, the results are shown in Fig.~\ref{fig:bbarrapidty} right panel. The agreement between data and FONLL calculation is slightly better at \s = 5.02 \TeV than at \s = 13 \TeV. The production cross section of \bbbar pairs per unit of rapidity at midrapidity ($\lvert$y$\lvert$ $<$ 4.5) is evaluated independently for the non-prompt \dzero and \jpsi. It is not needed to apply \pt extrapolation procedures to extend the \pt coverage down to 0 for the ALICE measured \dzero. The absence of low \pt cross sections of the non-prompt unmeasured \dzero is corrected by unfolding procedures, while the non-prompt \jpsi is measured down to \pt = 0. Eventually, the \bbbar pair production cross sections are evaluated from 0 and to 36 \GeVc, and the contribution from \pt greater than 36 \GeVc is much less than 1\%, which is negligible to the total \bbbar cross section. The results from two channels are combined as a weighted average according to their uncertainties, as two measurements are provided by two collaborations. The uncertainties can be considered independent, so the total uncertainties are used in the weight factor calculation, as shown by Eq.~\ref{eq:averageD} and~\ref{eq:averageDuncer}.
 
\begin{figure}[!h]
\begin{center}
 \includegraphics[width = 7cm]{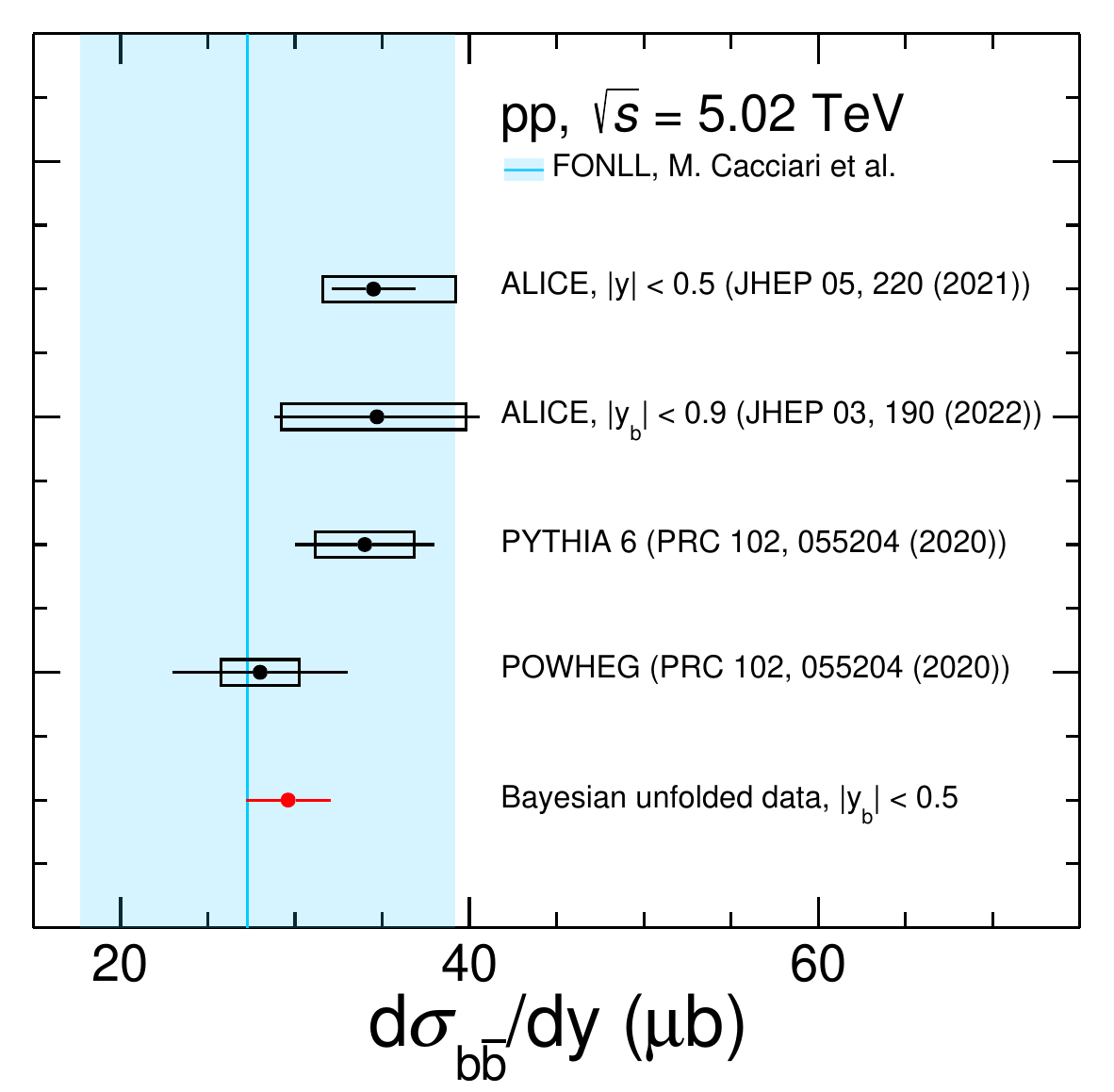}
 \includegraphics[width = 7cm]{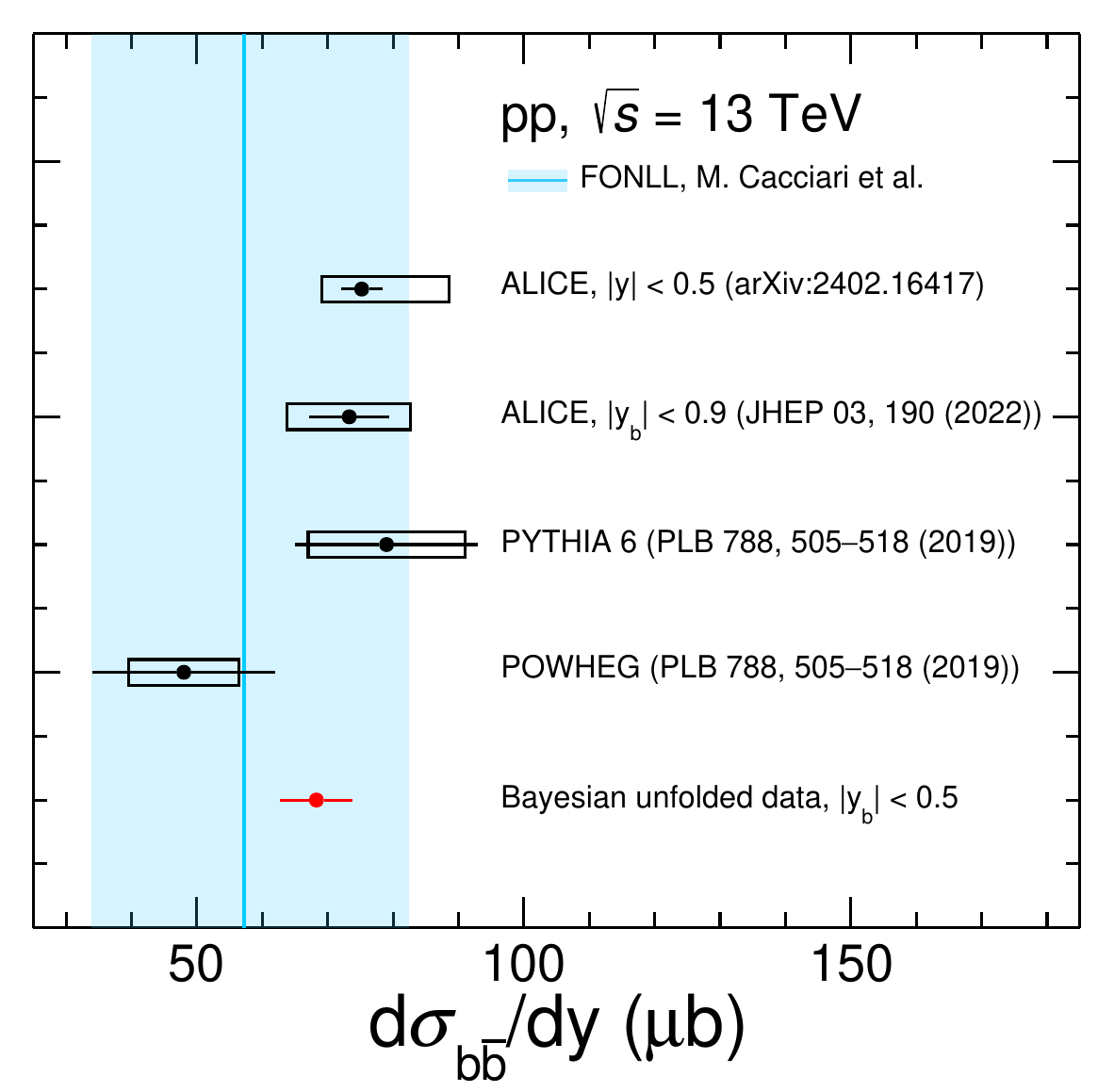}
\caption{Weight average \bbbar production cross section in \pp collisions at \s = 5.02 \TeV (left) and \s = 13 \TeV (right) at midrapidty, the results are compared with the existing measurements from ALICE collaboration~\cite{ALICE:2021mgk,ALICE:2024xln,ALICE:2018gev,ALICE:2020mfy,ALICE:2021edd} and FONLL predictions.}  
\label{fig:totalbbmidy}
\end{center}
\end{figure}

The total \bbbar production cross section at midrapidity is obtained in \pp collision at \s = 5.02 \TeV and \s = 13 \TeV. The correlated and uncorrelated uncertainties are propagated accordingly, following the same procedure as \dsigmady calculation described previously in this section. The final results of the total \bbbar production cross section at midrapidity are presented in Fig.~\ref{fig:totalbbmidy} for \s = 5.02 \TeV and \s = 13 \TeV. The new results are compared with the worldwide results provided by ALICE~\cite{ALICE:2021mgk,ALICE:2021edd,ALICE:2020mfy,ALICE:2018gev}. Furthermore, the new results are compared to the FONLL predictions, which are shown both for \s = 5.02 \TeV and \s = 13 \TeV. The new \bbbar cross section at midrapidity from the unfolding is compatible with the reported worldwide results by ALICE, as well as with PYTHIA and POWHEG~\cite{Sjostrand:2006za,Sjostrand:2014zea,Frixione:2007nw}. The uncertainties have been significantly improved compared to the worldwide presented results. The new results also agree with the FONLL predictions within their large uncertainties. The new \bbbar production cross section can be expected to provide constraints on the FONLL calculations, helping to further reduce the corresponding uncertainties of the FONLL predictions.

\begin{figure}[!htb]
\begin{center}
 \includegraphics[width = 10cm]{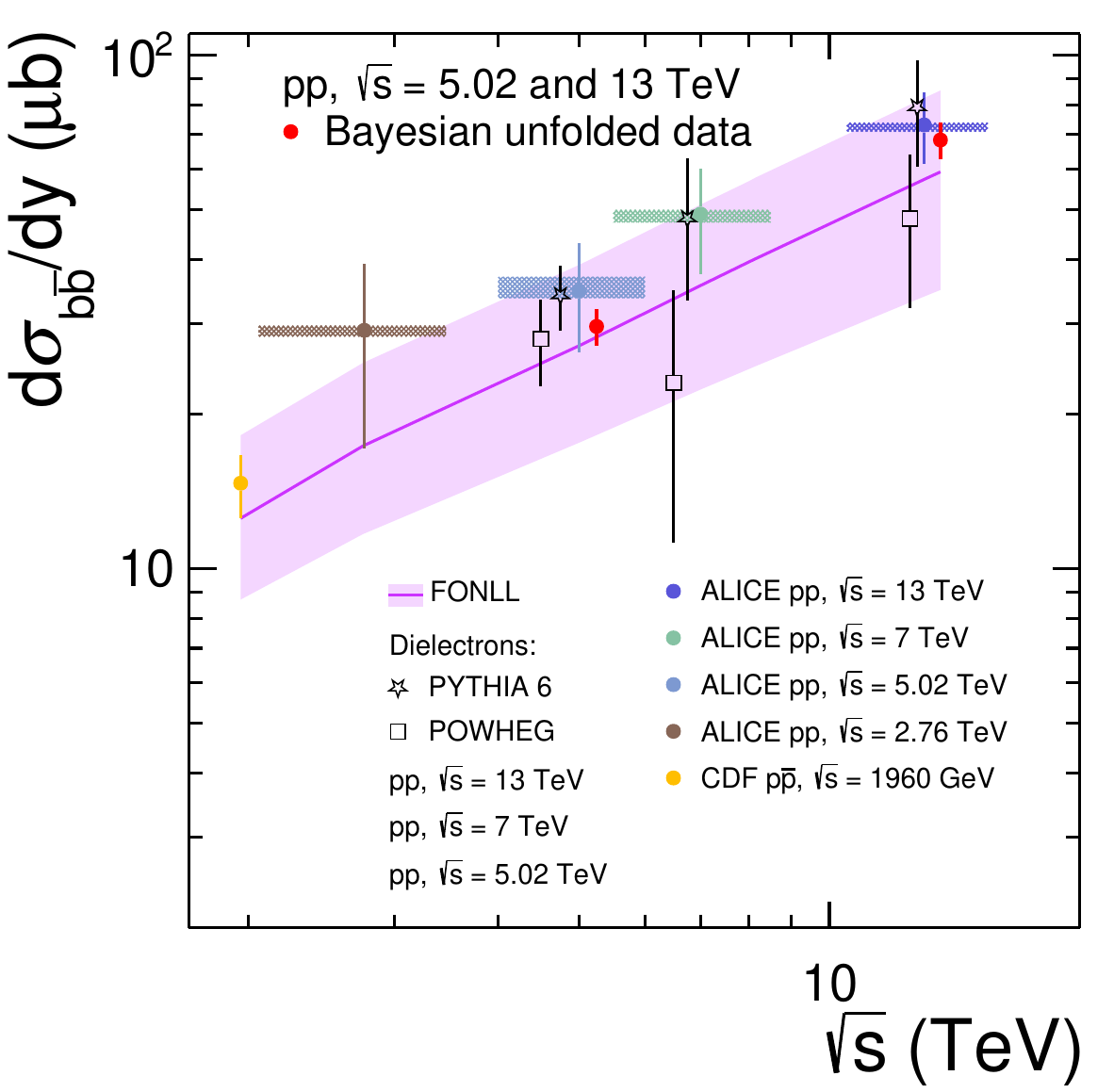}
\caption{Total \bbbar production cross section as a function of centre-of-mass energies at midrapidity in \pp collisions, the results are compared with the worldwide measurements~\cite{ALICE:2012acz,ALICE:2021edd,ALICE:2021mgk,ALICE:2014aev,CDF:2004jtw,ALICE:2018gev,ALICE:2018fvj,ALICE:2020mfy} and the FONLL calculations. }  
\label{fig:bbVsenergy}
\end{center}
\end{figure}

The total \bbbar cross section at midrapidity in \pp collisions at \s = 5.02 \TeV and \s = 13 \TeV are shown as a function of centre-of-mass energy in Fig.~\ref{fig:bbVsenergy}, together with other experimental measurements in \pp collisions from 1.96  up to 13 \TeV. The results are provided by ALICE~\cite{ALICE:2012acz,ALICE:2021edd,ALICE:2021mgk,ALICE:2014aev} and similar results in \ppbar collisions from CDF~\cite{CDF:2004jtw}. The measured results from dielectron by ALICE collaboration~\cite{ALICE:2018gev,ALICE:2018fvj,ALICE:2020mfy}, either PYTHIA or POWHEG simulations~\cite{Sjostrand:2006za,Sjostrand:2014zea,Frixione:2007nw}, are added as well for comparison. The FONLL calculations can describe the data within its large uncertainties,  despite the experimental points sitting on the upper side of the FONLL calculations. The \bbbar production cross section \dsigmady at midrapidity are:
$${\frac{\mathrm{d} \sigma_{\mathrm{\bbbar}}}{\mathrm{d} y}}_{\left|y_{\mathrm{b}}\right|<0.5}^{\sqrt{s}=5.02 \mathrm{TeV}}= \rm 29.6 \pm \rm 2.4 \,\, \rm \mu b,$$
$${\frac{\mathrm{d} \sigma_{\mathrm{\bbbar}}}{\mathrm{d} y}}_{\left|y_{\mathrm{b}}\right|<0.5}^{\sqrt{s}=13 \mathrm{TeV}}= \rm 68.3 \pm \rm 5.5 \,\, \rm \mu b$$
in \pp collisions at \s = 5.02 \TeV and \s = 13 \TeV, respectively. The uncertainties, including the statistical and systematical uncertainties from the measured non-prompt \dzero and \jpsi, are propagated separately. The statistical uncertainties are considered independent over \pt, while the systematic uncertainties including the global uncertainty are considered fully correlated over \pt. This extreme assumption overestimates the measured non-prompt systematical uncertainties. Further discussion on the total uncertainties can be found in the previous section~\ref{SystmaticalUncertantanties}. 

The total \bbbar production cross section $\sigma_{\mathrm{\bbbar}}$ in the full phase space in \pp collisions at \s = 5.02 \TeV and \s = 13 \TeV are: 

$${\sigma_{\mathrm{\bbbar}}^{\sqrt{s}=5.02 \mathrm{TeV}}} = \rm 191.3 \pm \rm 16.2 \,\, \rm \mu b,$$
$${\sigma_{\mathrm{\bbbar}}^{\sqrt{s}=13 \mathrm{TeV}}} = \rm 499.8 \pm \rm 39.6 \,\, \rm \mu b$$

The full phase space 4$\rm \pi$ is simulated by the unfolding matrix without any \y or \pt extrapolation. The propagation procedures of the uncertainties are the same as those for the \bbbar cross section at midrapidity \dsigmady evaluations. The new results of the total beauty production cross section in 4$\rm \pi$ full phase space are found to be consistent with the combined results of the LHCb and ALICE non-prompt \jpsi via the extrapolation approach~\cite{ALICE:2021edd}, where the full phase space extrapolations factor $\alpha_{4\rm \pi}$ is extracted from the FONLL calculations.

\FloatBarrier

\section{Conclusions}
\label{Conclusions} 

The \bbbar production cross sections in \pp collisions at centre-of-mass energy at \s = 5.02 \TeV and \s = 13 \TeV are obtained by using the Bayesian unfolding based on the measured non-prompt \dzero and non-prompt \jpsi mesons. The new results are compared with the FONLL predictions, as well as with existing experimental measurements.

The Bayesian unfolded open heavy hadron and \bbbar \pt-differential and integrated result obtained only from the ALICE non-prompt \dzero and only from the LHCb non-prompt \jpsi combined with the rapidity and \pt-shapes lead to mutually consistent results. Hence, this consistency provides critical cross-checks between different decay channels and measurement techniques. The weighted average results are computed according to the individual uncorrelated uncertainties. The FONLL prediction is in good agreement with Bayesian unfolding results within uncertainties in general, but the data is on the upper side of the FONLL calculation. 

The \pt-integrated \bbbar production cross sections \dsigmady in \pp collisions are presented at \s = 5.02 \TeV and \s = 13 \TeV. The cross section is significantly higher at midrapidity than at forward rapidity. The FONLL predictions are consistent within uncertainties to the unfolded data. The total \bbbar production cross section in the full phase space is compared with the existing data from ALICE measurements from the individual non-prompt \dzero and non-prompt \jpsi measurements at the midrapidity, which was evaluated by extrapolating \pt down to 0, and the large uncertainties are introduced by extrapolating process. 

The precision of the new results significantly improves the existing worldwide \bbbar production cross section. This will provide valuable validation and trigger improvements in pQCD-based calculations in \pp collisions. In addition, as the \bbbar production cross sections scale with the number of the binary collisions, it can be used as critical inputs for models of open beauty hadrons, as well as bottomonium-related calculations in heavy-ion collisions~\cite{Apolinario:2022vzg,Wu:2023djn}. Furthermore, these results can also serve as a good reference for future studies in heavy-ion physics. Finally, this study provides a show case for the usage of Bayesian unfolding relying on only partially reconstructed data from different experiments. The results are highly encouraging to employ the methodology to similar problems as the extraction of total \ccbar and \bbbar cross sections also in proton-nucleus, nucleus-nucleus and proton-proton collisions at different collision energies.

\FloatBarrier
\section*{Acknowledgments}
\label{Acknowledgments}

This work was supported by the National Natural Science Foundation of China with Grant No. 12061141008, No. 12375141, No.12105277 and No. 12205292, the Strategic Priority Research Program of Chinese Academy of Sciences with Grant No. XDB34030000. 

We would like to thank Anton Andronic, Andrea Dubla, Fiorella Fionda, Michael Winn, Fabrizio Grosa, and Gauthier Legras for the fruitful discussion and constructive suggestions for this work.

Xiaozhi Bai and Guangsheng Li contributed equally to this work as first co-authors.


\bibliographystyle{utphys} %
\bibliography{reference}

\end{document}